\documentclass[aip,reprint]{revtex4-1}

\usepackage{amsmath}
\usepackage{amssymb}
\usepackage{graphicx}
\usepackage{dcolumn}
\usepackage{bm}
\usepackage{multirow}
\usepackage{hyperref}
\usepackage{color}
\usepackage{xcolor}

\usepackage{appendix}

\newcommand{\rev}[1]{\textcolor{black}{#1}}
\newcommand{\revnew}[1]{\textcolor{black}{#1}}

\begin{document}

\title{Universality in diffusion-controlled nucleation and growth}

\author{Alexei V. Tkachenko}
\email{oleksiyt@bnl.gov}
\affiliation{Center for Functional Nanomaterials, Brookhaven National Laboratory, Upton, NY 11973, USA}

\begin{abstract}
Nucleation and growth are studied in a system that undergoes diffusion-controlled condensation under gradual changes in parameters, such as cooling.  It is demonstrated that when the Gibbs-Thompson effect becomes negligible, the system falls into a universal regime. i.e. the final droplet size distribution remains invariant under certain rescaling of system parameters. An approximate yet very accurate analytic form is obtained for the final droplet size distribution in this regime. 
\end{abstract}

\maketitle
\section{Introduction}
Nucleation and growth are fundamental processes in non-equilibrium statistical physics, playing a key role in condensed matter physics, chemistry, and materials science \cite{Becker1935,frenkel1946,Turnbull1949,Lifshitz1961,Wagner, Girshick1990,auer2001,kashchiev2000book,Kashchiev2003,agarwal2014,review2014}. These processes are critical to understanding phase separation, with applications in areas such as vapor condensation, multiphase segregation, colloidal aggregation, nanoparticle fabrication, and various biological processes \cite{talapin2001,Lee2016,review2014,bio2019}. Classical nucleation theory (CNT) explains the emergence of a new phase by formation of critical nuclei, whose size is determined by the balance between thermodynamic driving forces and surface free energy. Following nucleation, growth occurs as atoms or molecules from the parent phase diffuse and incorporate into the new phase.

Beyond CNT, the final stage of phase separation often involves Ostwald ripening, where larger droplets grow at the expense of smaller ones due to differences in chemical potential. Lifshitz-Slyozov theory provides insight into this coarsening process, predicting a self-similar size distribution and the scaling behavior of droplet radii \cite{Lifshitz1961,Wagner}. Additionally, inhomogeneous nucleation occurs at defects or interfaces, reducing the energy barrier for nucleation \cite{review2014}, while non-classical nucleation theories suggest deviations from CNT, including metastable intermediates and pre-nucleation clusters \cite{non-classical2011,Lee2016}. 

In this paper, we discuss a version of the problem  in which the system is subjected to a gradual change of parameters (typically, due to cooling) which leads to a widening gap in  chemical potential, $\Delta \mu$,   between the solution at the original concentration $c_0$ and the condensed state:
\begin{equation}
\frac{\Delta \mu(t)}{k_B T}=\ln \left(\frac{c_0}{c_s(t)}\right)=r(t-t_0) \label{eq:delta_mu}
\end{equation}
Here $c_s(t)$ represents the time-dependent saturation concentration of the solution of 
 of entities that we will refer to as  "molecules", even though they might be more complex building blocks. They undergo diffusion-controlled condensation into an aggregated phase, liquid, or solid. An important aspect of this version of the problem is that it lacks the Ostwald ripening regime. That is because the desorption process eventually becomes prohibitively slow and the overall size distribution of the condensate droplets becomes "frozen". Below, we focus on this final size distribution. 
 
\section{Nucleation and Growth Equations}
\rev{
The diffusion flux of molecules per unit surface area of a condensed droplet should equal the net adsorption flux. For a droplet $i$ of radius $R_i$ this can be expressed as
\begin{equation}
J_i=\frac{D\left(c-c(R_i)\right)}{R_i}=\frac{D}{\xi}\left[c(R_i)-c_s\exp\left(\frac{2\gamma v_0}{ k_B T R_i}\right)\right]
\end{equation}
Here $D$ is the diffusion coefficient of molecules, $c$ and \revnew{$c(R_i)$} are their concentrations far away and near the droplet surface, respectively,  $D/\xi$ is the adsorption rate, $v_0$ is the volume per molecule in the condensed phase, and  $\gamma$ is the surface tension of the droplet. The net adsorption flux includes the desorption term that depends on the chemical potential of the condensed phase. It is proportional to $c_s$ corrected by the exponential size-dependent factor due to the surface energy, i.e. the classical Gibbs-Thompson effect. \revnew{$c(R_i)$ can be eliminated from the above  equations as follows:
\begin{equation}
(R_i+\xi)J_i=D\left[c-c_s\exp\left(\frac{2\gamma v_0}{ k_B T R_i}\right)\right]
\end{equation}
This leads to}
\begin{equation}
J_i=\frac{D}{R_i+\xi}\left[c-c_s\exp\left(\frac{2\gamma v_0}{ k_B T R_i}\right)\right]
\end{equation}
The length scale $\xi$ depends on microscopic adsorption kinetics.  In this paper, we are interested in the diffusion-controlled regime \cite{kashchiev2000book,Kashchiev2003,agarwal2014} which corresponds to $R_i\gg \xi$ :
\begin{equation}
\dot{R}_i=J_iv_0=\frac{Dv_0}{R_i}\left[c(t)- c_s(t)\exp\left(\frac{2\gamma v_0}{ k_B T R_i}\right)\right] \label{eq:growth}
\end{equation}
At a given oversaturation level $S(t)=c(t)/c_s(t)$, a droplet starts growing once it exceeds the critical nucleus radius,
\begin{equation}
    R_{cr}=\frac{2\gamma v_0}{k_BT\ln S}
\end{equation}
}

\rev{According to the CNT, nucleation rate per unit volume has the following form \cite{Becker1935,frenkel1946,Turnbull1949,Lifshitz1961,Wagner, Girshick1990,auer2001,kashchiev2000book,Kashchiev2003,agarwal2014,review2014}:
\begin{equation}\dot{n}=Zfn_0e^{-\Omega} \label{eq:nucl0}
\end{equation}
Here  $f$ is the attachment rate of molecules to a critical nucleus. It should be noted out that in classical papers on CNT discussing condensation of vapor \cite{Becker1935,frenkel1946,Girshick1990}, this rate is assumed to be proportional to the thermal velocity of molecules and surface area. In light of our discussion, this corresponds to a reaction-limited regime with a length scale $\xi$ set by the mean free path of vapor molecules. This is justified as long as $R_{cr}\ll \xi$. The reaction-controlled regime is also a natural assumption for the nucleation within a condensed phase, such as solid in liquid\cite{auer2001}. In this paper, however, we are focused on diffusion-controlled regime\cite{kashchiev2000book,Kashchiev2003,agarwal2014} In this case, as follows from Eq. (\ref{eq:growth}), attachment rate is given by $f=4 \pi  R_{cr} c(t)$. 
}

\rev{Zeldovich factor {\cite{frenkel1946,Girshick1990,kashchiev2000book,auer2001, Kashchiev2003,agarwal2014}}, $Z=\frac{3v_0}{4\pi R_{cr}^3}\sqrt{\Omega/3\pi}$ in Eq. (\ref{eq:nucl0}), accounts for the fact that the size of a nucleus undergoes a random diffusion and only a small fraction of attachment events leads to an escape of the critical nucleus into the growth regime. Finally, 
$n_0e^{-\Omega}$ in Eq. (\ref{eq:nucl0}) is the  concentration of critical nuclei, which depends exponentially on the dimensionless nucleation barrier $\Omega$,
\begin{equation}
    \Omega(t) =\frac{4\pi \gamma R^2_{cr}}{3 k_B T} = \frac{16\pi \Gamma^3}{3 \ln^2 S} \label{eq:barrier}
\end{equation}
Here dimensionless surface tension has been introduced:
\begin{equation}
\Gamma  \equiv \frac{\gamma v^{2/3}_0}{k_BT}
\end{equation}
}
 
\rev{In the early days of CNT, the prefactor in front of the critical nucleus concentration, $n_0$, has been incorrectly identified with molecular concentration  $c$ \cite{Becker1935,frenkel1946}. However, this leads to an inconsistency with the mass action law, as the overall concentration of a critical nucleus containing $N_cr$ molecules should scale proportionally to $c^N_{cr}$, while that dependence is already included in the exponential Boltzmann factor $e^{-\Omega}$\cite{kashchiev2000book}. A common modification to CNT is to extrapolate the dependence of free energy on the nucleus size all the way to a single molecule, which leads to an estimate $n_0\simeq c_s e^{-{(36\pi)^{1/3}\Gamma}}$ \cite{Girshick1990,agarwal2014}. It was further argued in Ref. \cite{kashchiev2000book}, based on the Clapeyron-Clausius relationship, that this result corresponds to $n_0\simeq 1/v_0 $. It should be emphasized that the free energy of a critical nucleus in CNT is only approximated by two leading terms, bulk and surface. Therefore, any estimate of the prefactor $n_0$ should be taken with a grain of salt, as other effects contributing to the exponential factor are not fully accounted for, e.g. the nucleus shape fluctuation.  Fortunately, these corrections are not significant in the context of the present paper. We will therefore set $n_0=\chi/v_0$, where $\chi$ is a fudge factor potentially accounting for those additional effects.   After substituting the above results for $f$, $Z$, and $n_0$ into Eq. (\ref{eq:nucl0}), one obtains the nucleation rate in the following form \cite{kashchiev2000book, Kashchiev2003,agarwal2014}:
\begin{equation}
     \dot n = \frac{\chi Dc(t)\ln S(t)}{v_0^{2/3} \sqrt{\Gamma}}e^{-\Omega(t)}
    \label{eq:nucl}
\end{equation}  
}

\rev{
The overall mass conservation implies that the fraction of molecules still in solution, Let $\phi(t)=c(t)/c_0$   can be expressed as 
\begin{equation}
    \phi(t)=1-\frac{4\pi n(t)}{3\Phi_0}\langle R^3\rangle
    \label{eq:phi}
\end{equation}
Here $\Phi_0=v_0c_0$ is the overall volume fraction of molecules upon their condensation.
}

 \section{Scaling analysis and universal regime}
  Rate $r$ introduced in Eq.(\ref{eq:delta_mu}) sets the natural timescale in the problem. Accordingly,  Eq. (\ref{eq:growth}) suggests the following order-of-magnitude estimate of a typical droplet size: 
  \begin{equation}
  \label{lambda}
 \lambda_r=\sqrt{\frac{2D\Phi_0}{r}}
\end{equation}

\rev{
Our key approximation is that the droplets grow large enough for the Gibbs-Thompson effect to be sufficiently weak, i.e. $\lambda_r/v_0^{1/3}\gg \Gamma$. This allows one to neglect the size dependence of the desorption term in  Eq. (\ref{eq:growth}). For the sake of simplicity, in our further discussion, we will assume that nucleation primarily occurs at a sufficiently high level of oversaturation $S\gg 1$.  In practice, this corresponds to the limit of $\Gamma \gg 1$.  In this regime, the desorption term in Eq. (\ref{eq:growth}) can be neglected completely. Within this approximation, that equation can  be  as the growth of the dimensionless droplet radius $X_i\equiv R_i/\lambda_r$ in fictitious  dimensionless "time" $\tau=r\int\limits_0^t\phi(t')dt'$:
\begin{equation}
\frac{d X^2_i}{d\tau} \approx 1    \label{eq:growth1}
\end{equation} 
Note that the fraction of free molecules is related to the size distribution as $\phi=1-4\pi\nu \langle X^3\rangle/3$, where $\nu$ is  reduced droplet concentration:
\begin{equation}
    \nu=\frac{\lambda_r^3 n}{\Phi_0}
\end{equation}
}

\rev{
By  using Eq.(\ref{eq:barrier}) to express $\ln S$ in terms of nucleation barrier $\Omega$, Eq.(\ref{eq:nucl}) can be rewritten in the  following form:
\begin{equation}
\frac{d\nu}{d\tau} = \sqrt{\frac{4\pi }{3\Omega}}\Theta^{5/2}\Gamma e^{-\Omega}\label{eq:nucl_nu}
\end{equation}
Here, another key dimensionless parameter of the problem has been introduced:
\begin{equation}
     \Theta\equiv\frac{\chi^{2/5}\lambda^2_r}{v_0^{2/3}\Phi_0^{2/5}}= \frac{2\chi^{2/5}\Phi_0^{3/5} D}{v_0^{2/3}r}
\end{equation}
Since $\Theta$ is inversely proportional to rate $r$, it can be interpreted as a dimensionless cooling time.} 

Within our approximation of negligible Gibbs-Thompson Effect, the rescaled problem set by Eqs. (\ref{eq:growth1}) and (\ref{eq:nucl_nu}) has only two dimensionless parameters: $\Gamma$ and $\Theta$. In this  {\it universal regime}, the problem is invariant under the following rescaling, \rev {as it preserves $\Theta$, $\Gamma$ and $X_i=R_i/\lambda_r$:
\begin{align}
r/D &\rightarrow \eta r/D \\
\Phi_0 &\rightarrow \eta^{3/5} \Phi_0  \\
R &\rightarrow \eta^{-1/5} R 
\end{align}
}

\section{Approximate solution}

With time, the growing over-saturation leads to a rapid acceleration of the nucleation. This process is self-limiting: eventually, a sizable fraction of molecules end up in a condensed state. This results in the nucleation rate reaching its peak level and starting to decrease.  Without loss of generality, we choose $\tau=0$ around the time when the nucleation reaches its maximum rate.   
Let $\Omega^*=\Omega(0)$ \rev{and $S^*=S(0)$  be the corresponding values of the nucleation barrier, and supersaturation, respectively. We proceed by  expanding the nucleation barrier, Eq. (\ref{eq:barrier}) around its reference value $\Omega^*$:
\begin{equation}
   \Omega\approx\Omega^*-N_{cr}x+\frac{3N^2_{cr}}{4\Omega^*} x^2 
\end{equation}
Here $x=\ln S/S^*\approx\tau +\ln \phi \approx \tau +(1-\phi)$ and $N_{cr}$ is the number of molecules in a critical nucleus at the reference conditions:  
\begin{equation}
    N_{cr}=\frac{32\pi}{3}\left(\frac{\Gamma}{\ln S^*}\right)^3=\sqrt{\frac{3}{4\pi}}\left(\frac{\Omega^*}{\Gamma}\right)^{3/2}
\end{equation}
}

\rev{
After plugging the above expansion into Eq.(\ref{eq:nucl_nu}), one obtains
\begin{align}
\frac{d\nu}{dz} &= \sqrt{\frac{4\pi }{3\Omega^*}} \frac{\Theta^{5/2}\Gamma e^{-\Omega^*}}{N^*} F_\Omega(z)\label{eq:nucl1}
\\
F_\Omega(z)&\approx \exp\left(z-\psi(z)-\frac{3\left(z-\psi(z)\right)^2}{4\Omega^*}\right)\label{eq:g_x}
\end{align}
Here $z$ and $\psi(z)$ have the meaning of the fictitious time and the fraction of condensed phase, respectively, rescaled by factor $N^*$. The latter is close to the number of molecules in a critical nucleus, $N_{cr}$, up to a subdominant correction:
\begin{align}
  z&=N^*\tau=rN^*\int_{-\infty}^t \phi(t')dt'\\
  \psi(z)&=N^* \left(1-\phi(z)\right)\\
  N^*&=\sqrt{\frac{3}{4\pi}}\left(\frac{\Omega^*}{\Gamma}\right)^{3/2}\left(1+\frac{1}{2\Omega^*}\right)
    \label{eq:N_star}
\end{align}  
}


\rev{
As follows from Eq. (\ref{eq:growth1}), $X^2_i(\tau)\approx \left(\tau-\tau_i\right)=(z-z_i)/N^*$, where  $\tau_i=z_i/N^*$ corresponds to the moment of nucleation of the $i$-th droplet. 
This simple linear evolution of $X^2$ allows one to  relate the current  distribution function $f(X,z)$, to the nucleation dynamics:
\begin{equation}
    f(X,z)dX\approx d\nu(z')|_{z'=z-N^*X^{2}} \label{eq:f_x}
\end{equation}
Note that the distribution function is normalized as $\int  f(X,z)dX=\nu(z)$.
To complete the above set of equations, one needs to impose a self-consistency condition making sure that $\psi(z)/N^*$ is indeed the fraction of the condensed phase:
\begin{equation}
\frac{\psi(z)}{N^*}=\frac{4\pi }{3}\int\limits_0^\infty X^3 f(X,z)dX=\frac{4\pi }{3N^{*3/2}}\int\limits_{-\infty}^z (z-z')^{3/2} d\nu(z')
\end{equation}
}
By plugging in $d\nu/dz$ from Eq. (\ref{eq:nucl1}), one obtains the following integral relationship between functions  $\psi$ and $F_\Omega$ which are also related algebraically by Eq.(\ref{eq:g_x}):      
\begin{equation}   
   \psi(z) =\int\limits_{-\infty}^{z} (z-z')^{3/2}F_\Omega(z')dz' \label{eq:Psi}
\end{equation}
Here, the prefactor in front of the integral on the right-hand side is set to be $1$ to ensure that nucleation peaks close to $\tau=0$, i.e. $\psi(0)\sim 1$.  That leads to the following condition:
\rev{
\begin{equation}
1=\left(\frac{4 \pi}{3N^*}\right)^{3/2}\frac{\Theta^{5/2}\Gamma e^{-\Omega^*}}{\sqrt{\Omega^*}}  =\frac{\left(4\pi/3\right)^{9/4}\Theta^{5/2}\Gamma^{13/4} e^{-\Omega^*}}{\Omega^{*11/4}\left(1+\frac{1}{2\Omega^*}\right)^{3/2}} 
 \label{eq:unity}
\end{equation}
By taking log of this equation, one transforms it to  
\begin{align}
 \Omega^*&+\frac{11}{4}\ln \Omega^* +\frac{3}{4\Omega^*}\approx \widetilde{\Omega} \label{eq:w_eq}\\
 \widetilde{\Omega}&\equiv \frac{5}{2}\ln\Theta +\frac{13}{4} \ln \Gamma +\frac{9}{4}\ln\left(\frac{4\pi}{3}\right)\label{eq:tilde} 
\end{align} }
Here  $\widetilde{\Omega}$ is the zeroth-order approximation for $\Omega^*$ that depends logarithmically on the key parameters of the system: $\lambda_r/v_0^{1/3}$, $\Phi_0$, $\Gamma$, \rev{ as well as on the nucleation fudge factor $\chi$}. We found that the solution to Eq. (\ref{eq:w_eq}) can be very well approximated by the following explicit formula  valid for any physically relevant regime:  
\begin{equation}
    \Omega^* \approx \widetilde{\Omega}-3.3\ln \widetilde{\Omega} + 2.8 
\end{equation}

\begin{figure}
\centering
\includegraphics[width=.95\linewidth]{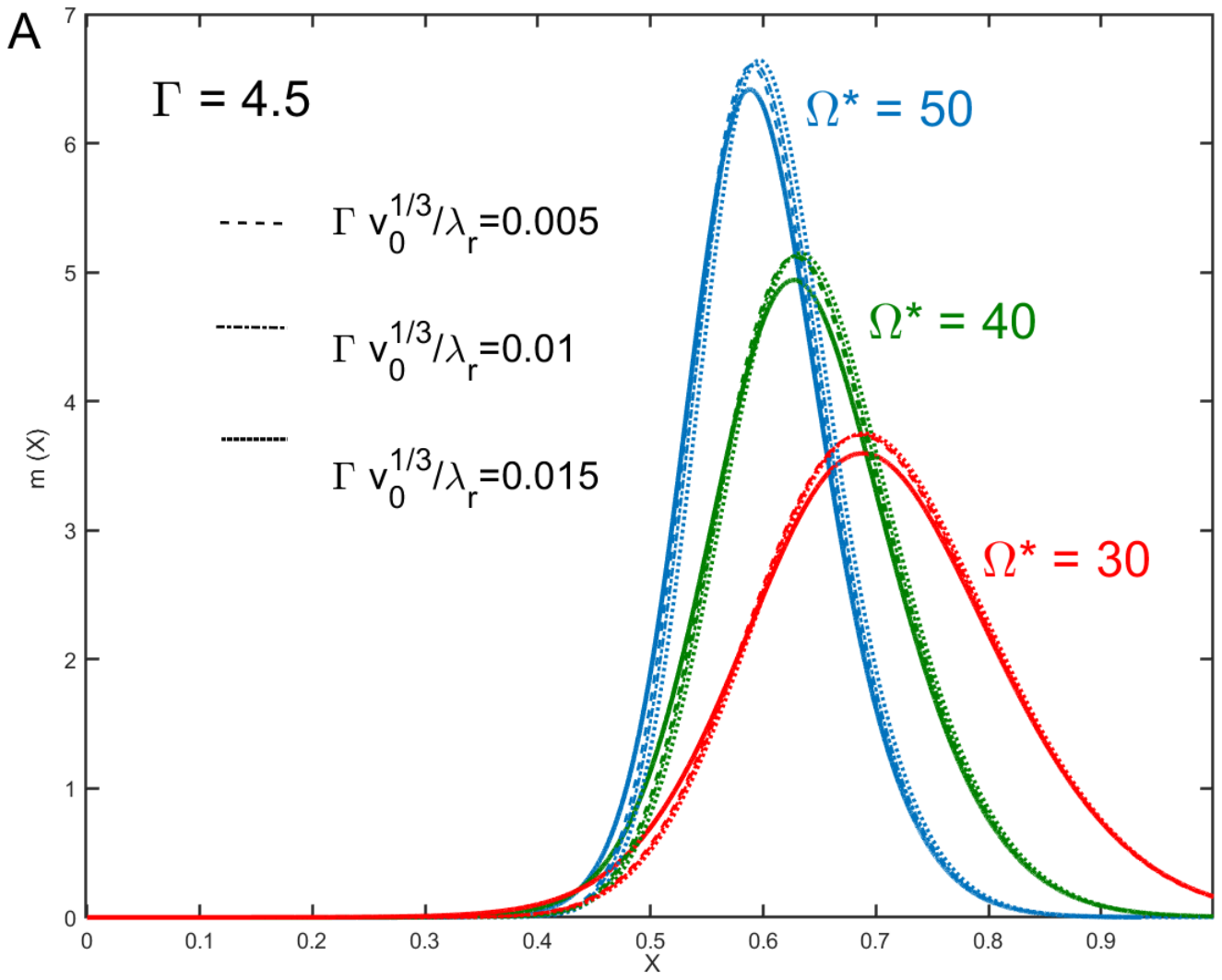}
\includegraphics[width=.95\linewidth]{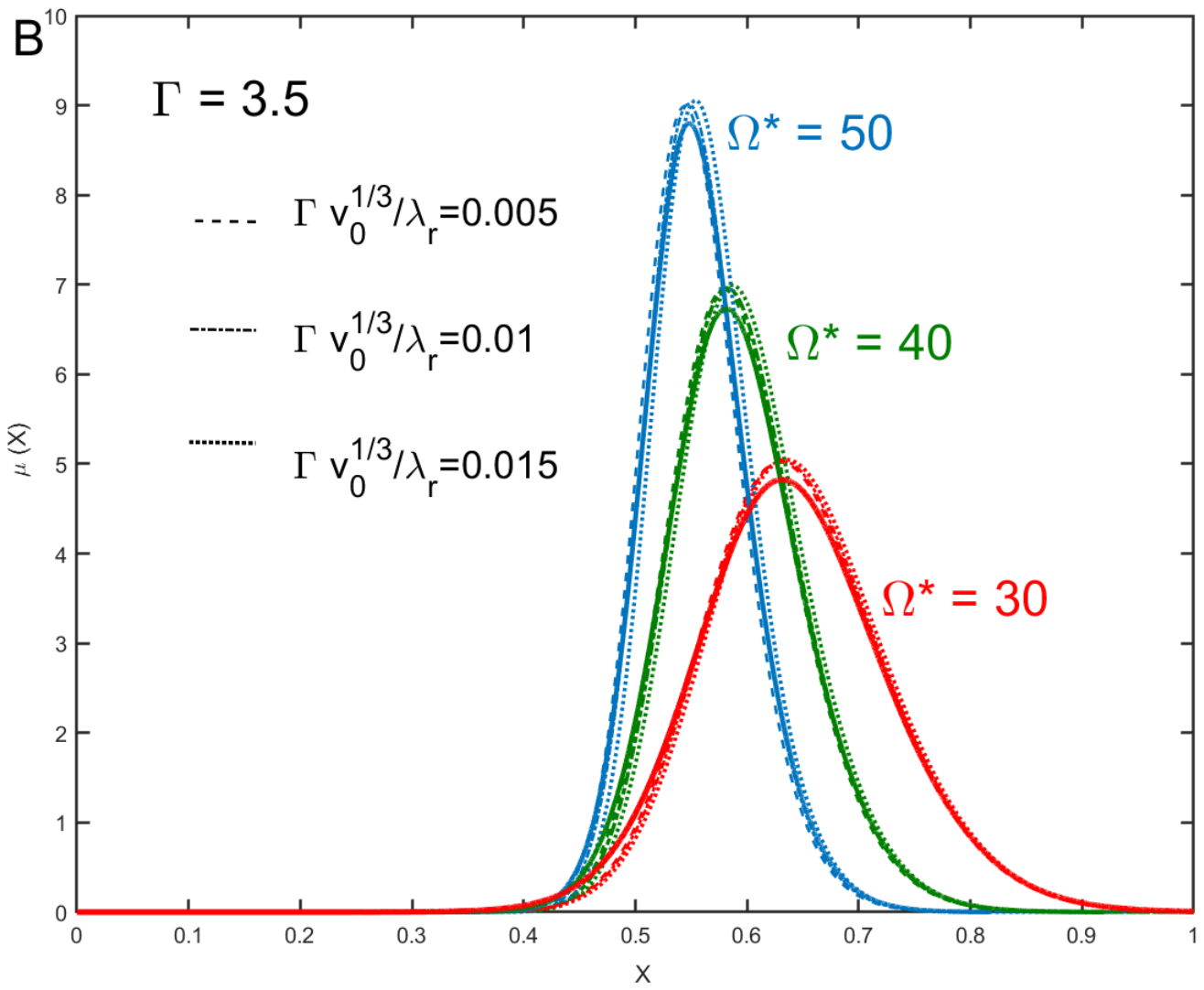}
\includegraphics[width=.95\linewidth]{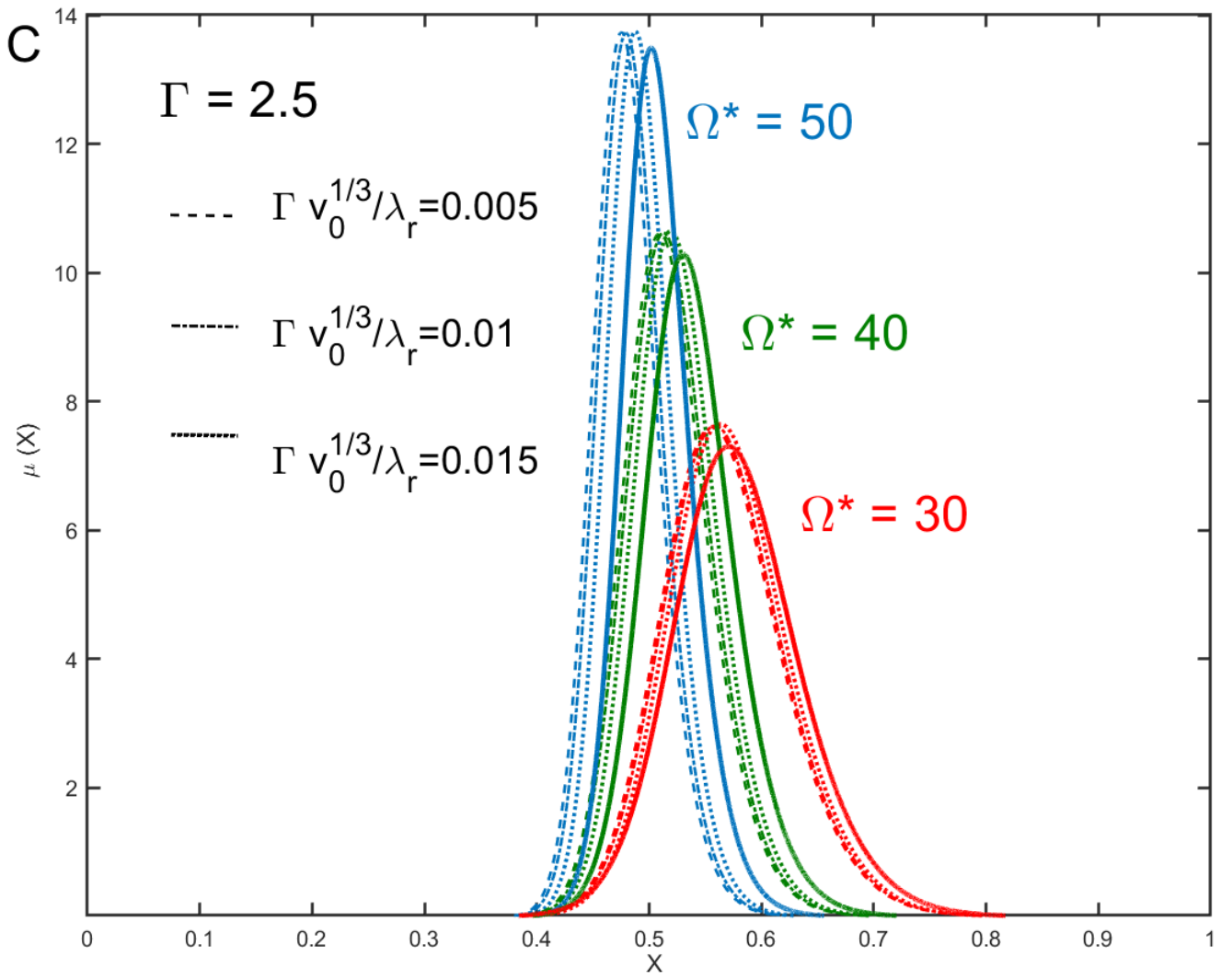}
\caption{Final mass distribution of the condensed phase over droplets of different size, $X=R/\lambda_r$. Solid lines represent the analytic results in the universal limit, Eq. (\ref{eq:f_R}. They are compared to the \rev{exact numerical solution of the nucleation and growth equations Eqs.(\ref{eq:growth}),(\ref{eq:nucl})}-(\ref{eq:phi}) with varying strength of Gibbs-Thompson effect characterized by dimensions parameter $\Gamma v_0^{1/3}/\lambda_r$. A, B, and C panels correspond to $\Gamma= 4.5, 3.5$, and $2.5$, respectively.}
\label{fig1}
\end{figure}

The growth of nucleated droplets continues until the solution gets completely depleted at finite value $\tau=\tau^*$ (but infinite physical time $t$). 
By combining Eqs. (\ref{eq:f_x}) and (\ref{eq:unity}) one obtains the  mass distribution reached at the end of the growth regime:   
\begin{align}
    m(X) &=\frac{4\pi X^3}{3}  f(X,z^*) =  \label{eq:f_R}\\
    &=2X^4N^{*3/2}F_\Omega\left(N^*\left[\frac{\alpha}{ N^{*1/3}}-X^2\right]\right)\nonumber
\end{align}
Here  dimensionless parameter $\alpha=z^* /N^{*2/3}$ is determined from normalization
condition:
\rev{\begin{equation}
    \int\limits_0^\infty m(X)dX=\frac{1}{N^*}\int\limits_{-\infty}^{z^*}\left(z^*-z \right)^{3/2} F_\Omega(z) dz=1
    \label{eq:alpha_eq}
\end{equation}}
Curiously, we found that all functions $\alpha(\Omega^*,N^*)$ for different values of $\Omega^*$ intersect in a single point, close to $\alpha \approx 0.8$ and $N^*\approx 5.85$. This observation allows one to construct a very accurate fit for $\alpha(\Omega^*,N^*)$ applicable everywhere in the physically relevant parameter range, $\Omega^*>20$:
\begin{equation}
\label{eq:alpha_approx}
\alpha \approx 0.801+0.114\cdot\left(1-\frac{10.4}{\Omega^*}+\frac{60}{\Omega^{*2}}\right)\left(1-\frac{5.85}{N^*}\right)
\end{equation}

\begin{figure}
\centering
\includegraphics[width=1\linewidth]{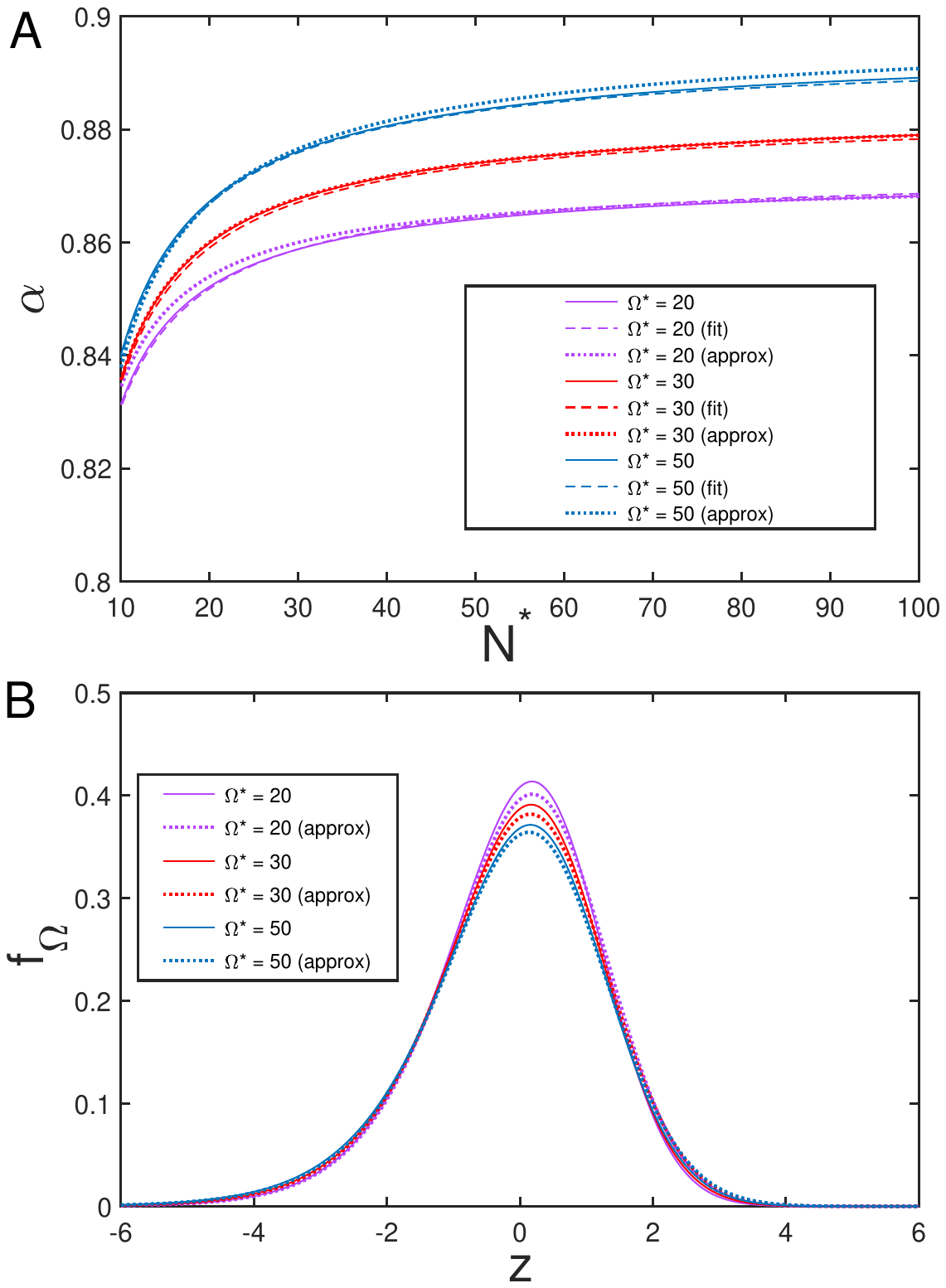}
\caption{Verification of the analytic approximations used. A: $\alpha$ vs $N^*$ for various values of $\Omega^*$. Solid lines represent the exact solution to Eq. (\ref{eq:alpha_eq}), dashed lines follow the fitted analytic formula  Eq. (\ref{eq:alpha_approx}), and dotted lines are obtained from Eq.(\ref{eq:f_omega_approx}), by using an approximate form of $F_\Omega$. B: Comparison of exact (solid) and approximate (dotted) shape of $F_\Omega (z)$.    }
\label{fig2}
\end{figure}

Within our approximation,  the final distribution Eq. (\ref{eq:f_R}) is determined by two parameters, $\Omega^*$ and $N^*$, which in turn are related to dimensionless combinations of system parameters $\Theta$ and $\Gamma$ through  Eqs. (\ref{eq:tilde})-(\ref{eq:w_eq}), and (\ref{eq:N_star}), respectively. \rev{Figure \ref{fig1} shows  our approximate result for $m(X)$  in the universal limit, compared with the numerical solutions of the original nucleation and growth equations: Eqs. (\ref{eq:growth}), (\ref{eq:nucl}), and (\ref{eq:phi}). While the key function $F_\Omega(z)$ 
can be determined numerically from Eqs. (\ref{eq:g_x}) and (\ref{eq:Psi}), we present below an approximate explicit formula that effectively captures its overall shape:}
\begin{align}
\label{eq:f_omega_approx} F_\Omega(z)&\approx\exp\left(\zeta(z)-\frac{\zeta(z)^2}{4\Omega^*}\right)\\
\zeta(z)&=z- \ln(1+e^z)\ln^\gamma(1+\beta e^z)\\
\beta &= \frac{11.8}{40/\Omega^*+1}\\
 \gamma&=0.5 + \frac{0.85}{\Omega^*}
\end{align}
The above approximations for $\alpha$ and $F_\Omega$ are shown in Figure \ref{fig2}, alongside the respective exact results. 

\section{Discussion and Conclusions}

\rev{
There are several ways in which the presented theory can be further generalized. In particular, it can be readily applied to systems where the nuclei and the growing particles are non-spherical, such as crystals growing out of a solution. In such cases, $R_i$ can represent any linear dimension of the condensed particle (e.g., edge size or hydrodynamic radius). Regardless of the specific choice, Eq. (\ref{eq:growth}) retains its form, with modifications only in a geometric prefactor $\zeta$ and the exponential term, which remains irrelevant within our approximation.
The dimensionless prefactor $\zeta$ can be determined for any particle shape by using the analogy between constant diffusion and electrostatics:  
\begin{equation}
\zeta=\frac{4\pi R^2 C_R}{3V_R}
\end{equation}
Here, $V_R$ is the volume of the particle with linear size $R$ and $C_R$ is the electrostatic capacitance of its shape expressed in units of length. This shape effect can be incorporated through an appropriate rescaling of the fundamental length scale: $\lambda_r\rightarrow\sqrt{\zeta} \lambda_r$. 
Another shape-related correction involves the relationship between $N^*$ (or $N_{cr}$), and $\Omega^*$. Specifically, an additional factor $6\sqrt\pi V_R/A_R^{3/2}$  must be included in Eq. (\ref{eq:N_star}) to account for the non-spherical relationship between volume $V_R$ and surface area $A_R$.  For example, in the case of diffusion-controlled growth of cubic crystals, this factor would be 
 $\sqrt{\pi/6}\approx 0.72$, and, assuming $R_i$ representing the edge size of a cube, $\zeta \approx 0.66$\cite{cube1997jcp}. Other geometric effects introduce only minor logarithmic corrections to $\Omega^*$, which can be absorbed into the fudge factor $\chi$.}
 
 \rev{Another avenue for generalization involves extending the results beyond diffusion-controlled nucleation and growth. This can be achieved by relaxing the approximation  $R_i\gg \xi$. In fact,  for $R_{cr}\ll \xi \ll \lambda_r$  a hybrid regime emerges, characterized by reaction-controlled nucleation coupled with diffusion-limited growth. In this case, the theory remains largely applicable with suitable corrections to the expression for $\Omega^*$, Eq. \ref{eq:w_eq}. Only in the regime $\xi \gtrsim \lambda_r$ would a substantial revision of the theory be required. }
 
In conclusion, we analyzed the classical nucleation and growth in the system with diffusion-controlled condensation,  under gradual change of parameters (such as cooling). In this case, there is no late-stage Ostwald ripening. Instead, the droplets of the condensed phase reach a steady state distribution, with characteristic  size given by $\lambda_r=\sqrt{\frac{2D\Phi_0}{r}}$, Eq. (\ref{lambda}). Furthermore, if that size is large enough for the Gibbs-Thompson effects to be negligible, the problem has a universal regime, with possibility of simultaneous rescaling of cooling rate $r$, volume fraction $\Phi_0$, and droplet size $R$. In this regime, the overall shape of the droplet size distribution is given by curves that depend on two dimensionless combinations of system parameters, $\Theta$ and $\Gamma$,  that characterize the cooling time and surface tension, respectively. 
We were able to obtain approximate but very accurate  expressions for those distribution functions.

These results are of potential importance for a wide range of applications in material science, condensed matter, and nanotechnology.

\begin{acknowledgments}
This research was done at and used resources of the Center for Functional Nanomaterials, which is a U.S. DOE Office of Science User Facility, at Brookhaven National Laboratory under Contract No. DE-SC0012704.
\end{acknowledgments}
\bibliography{main}

\end{document}